\documentclass[twocolumn]{aastex631}

\usepackage{amsmath}
\usepackage[mathlines]{lineno}

\begin{document}

\title{Dynamics of Star Cluster Formation: Mergers in Gas Rich Environments}

\author[0000-0003-3328-329X]{Jeremy Karam}
\affiliation{Department of Physics \& Astronomy, McMaster University, 1280 Main Street West, Hamilton ON, L8S 4M1, CANADA}

\author[0000-0003-3551-5090]{Alison Sills}
\affiliation{Department of Physics \& Astronomy, McMaster University, 1280 Main Street West, Hamilton ON, L8S 4M1, CANADA}







\begin{abstract}


We perform high resolution simulations of forming star clusters as they merge inside giant molecular clouds (GMCs) using  hydrodynamics coupled to N-body dynamics to simultaneously model both the gas and stars. We zoom in to previously run GMC simulations and resolve clusters into their stellar and gas components while including the surrounding GMC environment. We find that GMC gas is important in facilitating the growth of clusters in their embedded phase by promoting cluster mergers. Mergers induce asymmetric expansion of the stellar component of the clusters in our simulations. As well, mergers induce angular momentum in the clusters' stellar and gas components. We find that mergers can lead to an increase in the amount of dense gas present in clusters if a background gas distribution is present. We predict that this can lead to new star formation that can change the overall distribution of cluster stars in velocity space. Our results suggest that subcluster mergers in the presence of background gas can imprint dynamical signatures that can be used to constrain cluster formation histories.

\end{abstract}



\section{Introduction} \label{sec:intro}

The first few Myr of star cluster formation takes place embedded inside giant molecular clouds (GMCs) (\citealt{ladalada}). Simulations of this phase of star cluster formation have found that clusters form hierarchically, through mergers of smaller subclusters (e.g.\citealt{vazquez}, \citealt{Howard2018}, \citealt{chen}, \citealt{dobbs_1}, \citealt{rieder21}). Recent observations from JWST have explored hierarchical buildup in the context of young massive cluster formation (e.g. \citealt{fahrion}). Mergers, alongside interaction with surrounding GMC gas, have been shown to affect cluster size, mass, and density distribution (\citealt{karam}, \citealt{karam_23}, hereafter Paper I and Paper II respectively).

The dynamics of stars in young clusters are important observables used to probe their formation history. One example is cluster expansion which has been observed in many young star clusters (e.g. \citealt{kuhn_2019}, \citealt{kuhn_trifid}, \citealt{kounkel}, \citealt{della_croce_expansion}), and associations (e.g. \citealt{lim}). Recent work has found that expansion can be anisotropic around young clusters (e.g. \citealt{wright_2019}, \citealt{wright_2}) implying that the cluster structure before expansion may not be spherical. Further evidence of substructure has been inferred from observation of the distribution of cluster stars in velocity space. \citet{wright} find evidence of kinematic substructure inside components of the Sco Cen OB association, and use this to argue that the distribution of stars in position space was once substructured. 

Another potential dynamical imprint of the embedded phase of star cluster formation is cluster rotation. Low rotational velocities have been observed in globular clusters (e.g. \citealt{bellazzini}, \citealt{paolo_1}) and in young massive star clusters (e.g. \citealt{r136_rotation}). If young massive clusters are present day analogues of globular clusters (see \citealt{zwart}), then globular cluster rotation may be inherited from the evolution of young clusters inside their embedded clouds. A full understanding of the evolution of embedded clusters must factor in these dynamical observations.

Simulations of young star cluster formation and evolution have been able to reproduce some of these observed dynamical signatures. N-body simulations (\citealt{geyer_burkert}, \citealt{baumgardt_kroupa}), along with simulations that include N-body, hydrodynamics, and stellar feedback (e.g. \citealt{pelupessy}), have shown that cluster expansion can be caused by the expulsion of the gas within which young clusters are embedded. These simulations, however, assume that the embedded phase of star cluster formation results in a cluster that is spherical immediately before gas expulsion. To better understand the hierarchical buildup of the cluster in the embedded phase, one can turn to simulations of cluster evolution inside GMCs. Such simulations have found that star clusters can form with primordial rotation and that the strength of their rotation may depend on their mass (\citealt{lahen}, \citealt{chen}). 

Large scale simulations like these come with their own set of limitations in the form of resolution. If GMCs are very massive (10$^{6-7}$M$_\odot$), one cannot resolve the evolution of the individual stars present in the GMC in a simulation while including all physical mechanisms that affect the GMC (e.g. \citealt{Howard2018}, \citealt{ali}). Instead, such large scale simulations employ the use of sink particles to model their clusters. A standard procedure is to use the sink particle implementation described in \citet{sink} where the cluster is modelled as a point particle with an accretion radius. The sink carries with it parameters that describe the cluster it represents (i.e. its total mass, mass in stars, position, and velocity).

In this work, we take regions from the \citet{Howard2018} (hereafter H18) GMC simulations and isolate them so that we can resolve the sink particles as collections of stars and gas that are more representative of a young, embedded cluster. We examine how the evolution inside the GMC environment affects the stars and gas present in the clusters through the inclusion of realistic background gas from the H18 GMC simulation. We also perform simulations without background gas that build upon our work in Paper I by investigating the effects of geometry on cluster mergers.

In Section \ref{sec:methods} we discuss the initial conditions of our simulations and how we convert the H18 data to our higher resolution framework. In section \ref{sec:stars} we discuss the evolution of isolated star cluster mergers with varying impact parameters. In Section \ref{sec:stars_corey} we discuss the dynamics of the stellar component of the clusters in our zoom-in simulations. In Section \ref{sec:gas} we discuss the evolution of the gas component of our zoom-in simulations. In Section \ref{sec:discussion}, we summarize and discuss the implications of our results and how they relate to observations of older star clusters.

\begin{center}
\begin{table*}
\hspace{-1cm}
\begin{tabular}{ccc|cccc}

  Simulation Name &  M$_{\mathrm{BG,g}}$ [10$^4$M$_\odot$] 
  
  & b [L$_{\mathrm{50,MM}}$]
  & Cluster Name & M$_\mathrm{s}$ [10$^3$M$_\odot$] & M$_\mathrm{g}$ [10$^4$M$_\odot$] & L$_\mathrm{50}$ [pc] \\
  \hline 

   \texttt{b0p5} & 0.0 & 0.5 & A & 0.2 & 0.2 & 0.4 \\
    & & & B & 0.2 & 0.4 & 0.4 \\
    \hline
    \texttt{b1} & 0.0 & 1.0 & A & 0.2 & 0.2 & 0.4 \\
    & & & B & 0.2 & 0.4 & 0.4 \\
    \hline
    \texttt{b1p5} & 0.0 & 1.5 & A & 0.2 & 0.2 & 0.4 \\
    & & & B & 0.2 & 0.4 & 0.4 \\
    \hline
    \texttt{b2} & 0.0 & 2.0 & A & 0.2 & 0.2 & 0.4 \\
    & & & B & 0.2 & 0.4 & 0.4 \\
    \hline
    \hline
  
   &  & &  A & 0.2 & 0.2 & 0.4\\
   \texttt{region1} & 4.9 
   
   & 1.3
   &  B & 0.2 & 0.4 & 0.4\\
   &   & & C & 0.6 & 0.5 & 0.5\\
  \hline
   &    & & A & 0.7 & 0.8 & 0.7\\
  \texttt{region2} & 7.5 
  
  & 1.9
  & B & 0.9 & 0.6 & 0.6\\
   &   & & C & 6.5 & 2.8 & 0.9\\
   \hline
   &    & & A & 0.6 & 0.4 & 0.6\\
  \texttt{region3} & 5.0 
  
  & (1.1, 0.7)
  & B & 0.7 & 0.1 & 0.3\\
   &   & & C & 9.1 & 1.0 & 0.8\\
  
\end{tabular}
\caption{Parameters for our simulations. Column 1: the simulation name, column 2: the mass of background gas, column 3: the impact parameter of the merger(s) in units of the stellar component 50\% Lagrangian radius of the cluster with the more massive stellar component in the merger, column 4: the star cluster name, column 5: the initial stellar mass of the cluster, column 6: the initial gas mass of the cluster, column 7: the initial stellar 50\% Lagrangian radius of the cluster.}
\label{tab:regions}
\end{table*}
\end{center}

\section{Methods}
\label{sec:methods}

\subsection{Numerical Methods}
\label{sec:sinks}
We perform our simulations using the Astrophysical Multipurpose Software Environment (AMUSE) (\citealt{zwart2009}, \citealt{amuse_3}, \citealt{amuse_2}). AMUSE contains codes that evolve the equations of gravity and hydrodynamics. As well, it allows for the communication between codes meaning that we can simulate the evolution of the stellar and gas components in a cluster and analyze how both components interact with each other. We use \texttt{hermite0} (\citealt{hermite}) as our N-body code, and \texttt{GADGET-2} (\citealt{gadget2}) as our smoothed particle hydrodynamics (SPH) code. We use \texttt{BRIDGE} (\citealt{bridge}) as our communication scheme with \texttt{BHTree} (written by Jun Makino based on \citealt{tree}) as the connecting algorithm. For more details on our numerical scheme, see Paper I. 

We use Plummer (\citealt{plummer1911}) spheres of stars and gas to set up our star clusters. We take the total stellar and gas mass of the cluster from sinks in the H18 simulation. We assign the masses of the individual stars using a Kroupa (\citealt{kroupa2001}) IMF with a lower mass limit of 0.15M$_\odot$, and an upper mass limit of 100M$_\odot$. We give our SPH gas particles a mass of $m_{\mathrm{SPH}} = 0.03$M$_\odot$ (we discuss this choice in more detail in Section \ref{sec:flash_to_SPH}) and set the gas temperature to 10K. We use an adiabatic equation of state for the gas. We use the method outlined in \citet{aarseth} to sample the velocities of the stars and gas in the cluster, and scale the velocities such that our clusters are initially in virial equilibrium (2$K_{\mathrm{s,g}}/|U_{\mathrm{s,g}}| = 1$ where $K_{\mathrm{s,g}}$ is the kinetic energy of the stars or gas, and $U_{\mathrm{s,g}}$ is the potential energy of the stars or gas).We chose the scale radius of both the stellar and gas Plummer spheres such that the density at the clusters half mass radius is consistent with young massive cluster observations shown in \citet{zwart} ($\rho_{\mathrm{hm}} \approx 10^3 - 10^4$M$_\odot$pc$^{-3}$). For more detail regarding the set up of our star clusters, see Paper I.

\subsection{Isolated Merger Simulations}
\label{sec:comp_sims}

The merger simulations presented in Paper I were all head-on collisions. This is not necessarily true for mergers in the H18 GMC simulations. To build upon the results from Paper I we first run off-axis cluster merger simulations. We perform simulations of two clusters merging along the x-axis with their centres offset along the y-axis by some impact parameter value that we label as $b$. The parameters for the clusters in these simulations can be seen in the top four rows of Table \ref{tab:regions}. The mergers in all four of these simulations have the same collisional velocity of 6.9kms$^{-1}$. The stellar and gas masses of these clusters, as well as the collision velocity, were chosen to be the same as the clusters that merge in the first of our more realistic merger simulations in a background gas potential, described in the next section.

\subsection{Clusters in Realistic Background Gas Distributions}
\label{sec:flash_to_SPH}

The bulk of our results come from our simulations of clusters evolving inside realistic background gas distributions. The initial conditions for the clusters and background gas in these simulations are taken from the H18 GMC simulations. These were a set of radiative, magneto-hydrodynamical simulations that used the \texttt{FLASH} (\citealt{FLASH}) adaptive mesh refinement (AMR) grid code to follow the formation of star clusters inside an evolving GMC.

We select three regions from their simulation of a 10$^7$M$_\odot$ GMC with a metallicity of 0.1Z$_\odot$ to use as our initial conditions. All three of these regions contain three sink particles each. In the H18 simulations, two sinks in \texttt{region1} will merge in the vicinity of the third, more massive sink. The same is true in \texttt{region2}. In \texttt{region3}, all three sinks merge until one sink is left by the end of the simulation. The widths along x, y, and z of \texttt{region1}, \texttt{region2}, and \texttt{region3} are (20, 10, 20)pc, (14, 20, 14)pc, (10, 20, 20)pc respectively. Once we selected our regions, we convert the background gas present in each region from the H18 grid  to an SPH particle distribution. We use a method adapted from \citet{ray-raposo} where the goal was to increase the resolution of SPH simulations by decreasing the SPH particle mass.

We begin by distributing $N_{\mathrm{SPH}} = M_{\mathrm{cell}}/m_{\mathrm{SPH}}$ (where $M_{\mathrm{cell}}$ is the mass of a given grid cell) SPH particles using a Gaussian around the centre of each grid cell. Each H18 cell is either 0.67 or 1.35pc on a side in these regions of the simulation, so there are approximately 14 thousand cells in \texttt{region1}, 8 thousand cells in \texttt{region2}, and 13 thousand cells in \texttt{region3}. We could choose to uniformly distribute the SPH particles inside each grid cell, but because the grid cells from the H18 simulations had large widths, the SPH representation would maintain the grid shape for much of the simulation. We choose the standard deviation of the Gaussian to be the width of the grid cell because this choice allows for the best conservation of density between the grid and SPH representations of the gas. We choose $m_{\mathrm{SPH}}$ = 0.03M$_\odot$ because this allows us to adequately sample SPH particles for a given cell without unnecessarily increasing the computation runtime. For \texttt{region1}, \texttt{region2}, and \texttt{region3}, we have a total of approximately 1.6 million, 2.5 million, and 1.7 million SPH particles in background gas respectively.

Next, we give all the SPH particles the same velocity as that of the cell that they are representing. To prevent the SPH representation from clumping, we apply a velocity dispersion of $0.5$kms$^{-1}$ to the velocity of the SPH particle inherited from its parent cell. We then divide the thermal energy of a given grid cell evenly amongst all the SPH particles we place around its centre such that $E_{\mathrm{th,SPH}}$ = $E_{\mathrm{th,cell}}$/$N_{\mathrm{SPH}}$ where $E_{\mathrm{th,SPH}}$ is the thermal energy of a given SPH particle, and $E_{\mathrm{th,cell}}$ is the specific thermal energy of the cell that the particles are representing. The background gas in our selected regions is not particularly uniform, but has a density range of about 4 orders of magnitude as can be seen in Figures \ref{fig:spur_snaps}, \ref{fig:r3_snaps}, and \ref{fig:jelly_snaps}.

We find that this method gives good overall agreement between the total mass in the grid and SPH representation for the regions of the H18 simulation we selected. The total mass of our SPH representation is within 0.1\% of the total mass from the grid simulation for all three regions we selected. As well, we find that the energies are well conserved between both representations. For all regions selected, the potential energy of the SPH representation is $\approx95$\% that of the original grid representation, the kinetic energy of the SPH representation is $<0.5$\% higher than the original grid representation, and the thermal energy of the SPH and grid representation are the same.

Once we have converted the background gas into an SPH representation, we replace the sink particles present in each region with star clusters made up of stars and gas using the method outlined in Section \ref{sec:sinks} and in Paper I. We show the parameters of these simulations in the bottom three rows of Table \ref{tab:regions}.

\subsection{Identifying Star Clusters}
\label{sec:dbscan}

To identify clusters throughout our simulations, we begin by using DBSCAN (\citealt{dbscan}, \citealt{dbscan2}) to spatially cluster the stars ensuring that there are five neighbouring stars within a given radius away from each star (\citealt{jorg}). We choose this radius such that the initial spatial clustering by DBSCAN returns the same number of clusters as there are sink particles initially in the region from H18. Once we have spatially clustered the stars, we check that each star is bound to its respective cluster by calculating its kinetic and potential energies. Finally, we check if spatially unclustered stars are bound. If they are, we assign it to the cluster with which it is most bound. 

To assign gas to each cluster, we first make sure that the gas is bound. We then calculate the potential energy felt on the gas from each cluster and assign each bound gas particle to the cluster to which it is most bound.

\section{Isolated Off-Axis Mergers}
\label{sec:stars}

We first analyze our off-axis isolated merger simulations. Lower impact parameter mergers only result in a slight increase in the amount of unbound stellar mass in the resultant cluster. By the end of simulations \texttt{b0p5}, \texttt{b1}, \texttt{b1p5}, and \texttt{b2}, the resultant cluster has lost of $14, 10, 7,$ and $5\%$ of its total stellar mass through the merger. Though the resultant clusters lose less gas mass than stellar mass in each simulation, we find the same trend with changing impact parameter regarding the bound gas component of the resultant cluster.

Simulations \texttt{b1}, \texttt{b1p5} and \texttt{b2} result in a cluster that is non-monolithic (two distinct stellar and gas components). In Paper I, we found non-monolithic resultant clusters for mergers with collisional velocities $\gtrapprox 10$kms$^{-1}$. The simulations presented here all have lower collisional velocities. Therefore, a larger impact parameter can lower the required collisional velocity to produce a non-monolithic cluster after a merger.

Though we do not include a star formation prescription in our simulations, we can track the amount of dense gas to predict how star formation would be affected. We consider gas with densities above $10^{3}$, $10^4$ and $10^5$cm$^{-3}$ as these have been considered as density thresholds above which star formation occurs in observations and simulations (e.g \citealt{elmegreen_2001}, \citealt{enoch_2008}, \citealt{lada2012}, \citealt{dobbs_1}). The amount of gas with densities above these thresholds peaks at the time of the merger, and decreases below its original value by the end of the simulation. Smaller impact parameter mergers result in a larger but shorter-lived increase in gas above 10$^4$ and 10$^5$cm$^{-3}$. Higher impact parameter mergers result in clusters which hold on to more gas above these density thresholds at the end of the simulation.

\begin{figure*}
\hspace{0.5cm}
\vspace{-0.2cm}
    \includegraphics[scale=0.28]{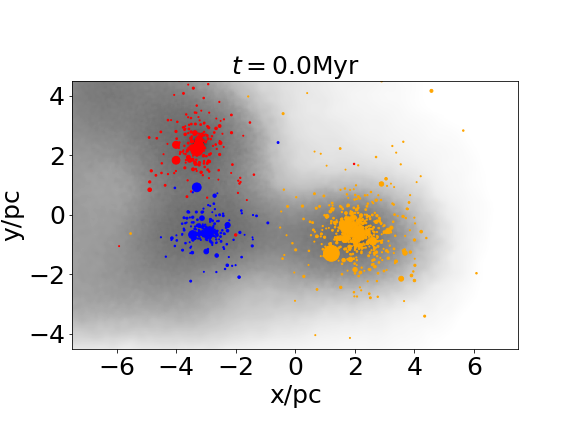}
    \includegraphics[scale=0.28]{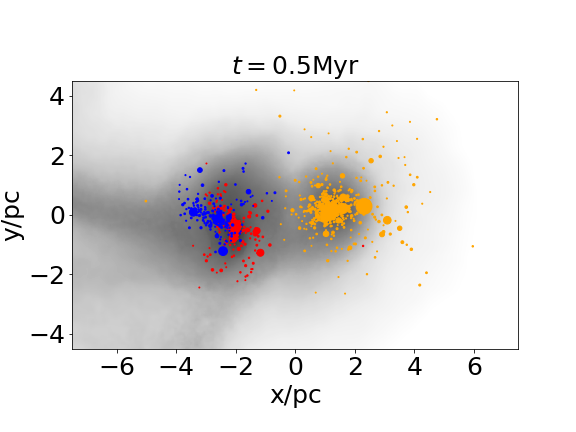}
    \includegraphics[scale=0.28]{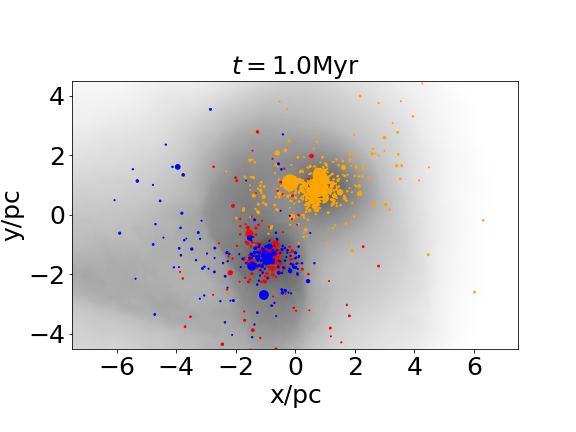}
    
    \hspace{1.3cm}
    \includegraphics[scale=0.28]{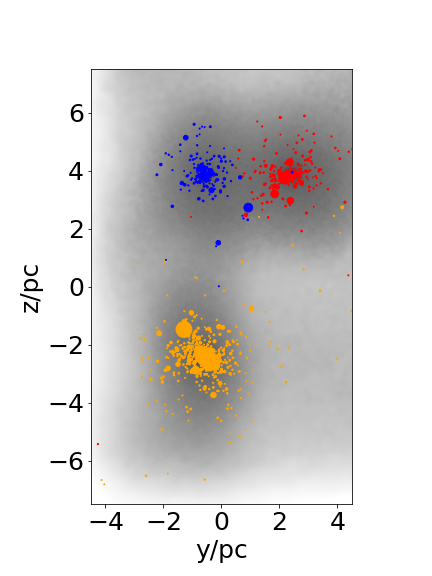}
    \hspace{1.3cm}
    \includegraphics[scale=0.28]{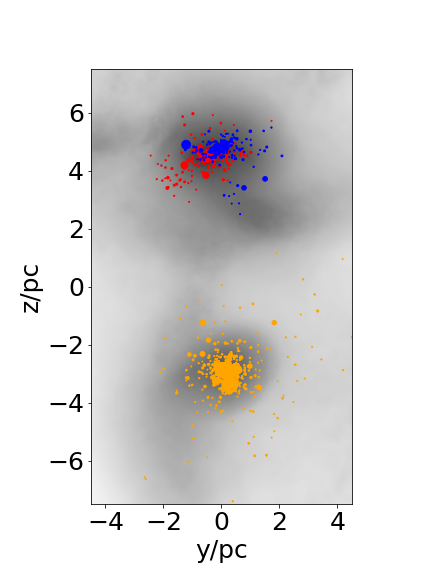}
    \hspace{1.3cm}
    \includegraphics[scale=0.28]{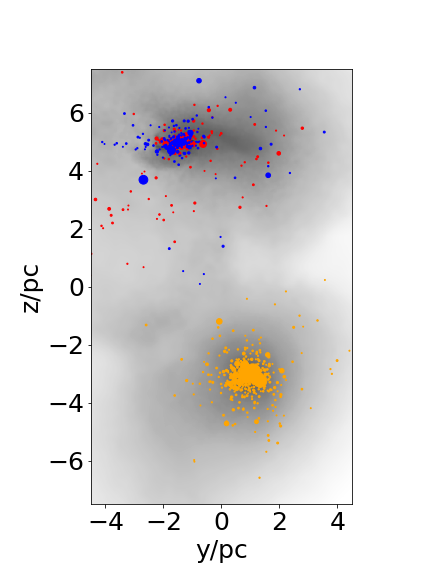}
    \caption{Snapshots of the stars and gas from the \texttt{region1} simulation in the x-y (top row) and y-z (bottom row) planes. The circles represent the stars in each cluster and their size scales with the stars mass. The lowest and highest mass stars in this simulation are 0.15M$_\odot$ and 41.70M$_\odot$ respectively. The gas is shown in black with darker regions showing gas with higher densities. The range in gas densities shown here is 0.1$-$10$^3 $M$_\odot$pc$^{-3}$.}
    \label{fig:spur_snaps}
\end{figure*}

\section{Zoom-In Regions: Stars}
\label{sec:stars_corey}

\subsection{Region1}

We show snapshots of a portion of the \texttt{region1} simulation in Figure \ref{fig:spur_snaps}. The circles are colour-coded depending on the cluster to which they initially belong with red, blue, and orange corresponding to clusters A, B, and C respectively. Initially, cluster A is moving towards cluster B in the x-y plane. At t$\approx$0.1Myr, our cluster identification scheme determines that the merger has begun. At t$\approx$0.4Myr, the stellar components of the two clusters reach their closest approach. They stay as a monolithic cluster for the remainder of the simulation. We call the resultant monolithic cluster ``cluster AB''. Cluster C does not merge with cluster A, cluster B or cluster AB throughout the simulation.

 The mass ratio, defined as the total (stellar and gas) mass of the more massive cluster divided by that of the less massive cluster, of the merger of cluster A with cluster B is $f_{\mathrm{M}} = 1.9$ and the collisional velocity of the merger is 6.9 kms$^{-1}$. The impact parameter of the merger of cluster A with cluster B is 1.3L$_{\mathrm{50,A}}$ where L$_{\mathrm{50,A}}$ is the 50\% Lagrangian radius of cluster A. In the previous section, we found that the simulation with similar properties (\texttt{b1}, and \texttt{b1p5}), but with no background gas resulted in a non-monolithic cluster. This demonstrates the importance of the background gas distribution on the structure of merging star clusters as it can be used to keep clusters monolithic. We confirmed the impact of the background gas on the evolution of the clusters in the \texttt{region1} simulation by running a simulation with the same setup as \texttt{region1}, but with no background gas distribution. We find that when the background gas is not included, clusters A and B do not merge to form a monolithic cluster by the end of the simulation.

The merger of cluster A and B in the \texttt{region1} simulation results in a small fraction of stars becoming unbound from cluster AB. The total mass in unbound stars in this simulation is $<$3\% of the total stellar mass which is less than the amount of unbound stellar mass produced from our merger simulations in Paper I and in Section \ref{sec:stars}. This is expected as the addition of background gas in these simulations provides an added potential that helps keep the stars and gas bound to their respective clusters.

\subsubsection{Velocity Space Distribution}
\label{sec:vel_dist_spur}
\begin{figure}
    \centering
    \includegraphics[scale=0.4]{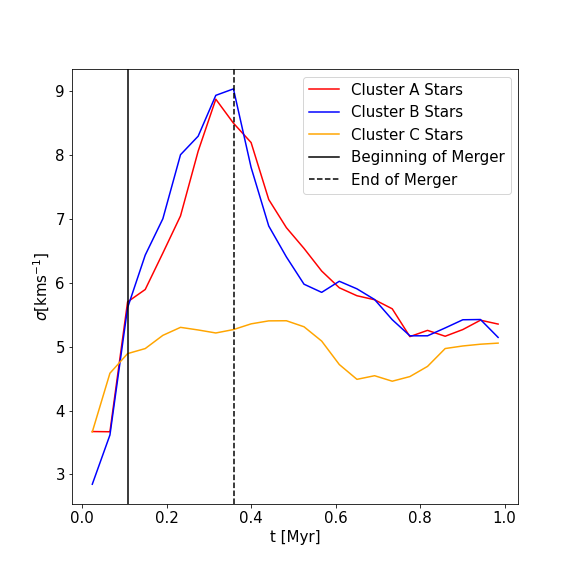}
    \caption{Velocity dispersion of the stars originally belonging to clusters A (red), B (blue), and C (orange) throughout the \texttt{region1} simulation. The solid and dashed black lines show the beginning and end of the merger of cluster A with cluster B respectively.}
    \label{fig:sigma1}
    
\end{figure}

\begin{figure}
    \centering
    \includegraphics[scale=0.27]{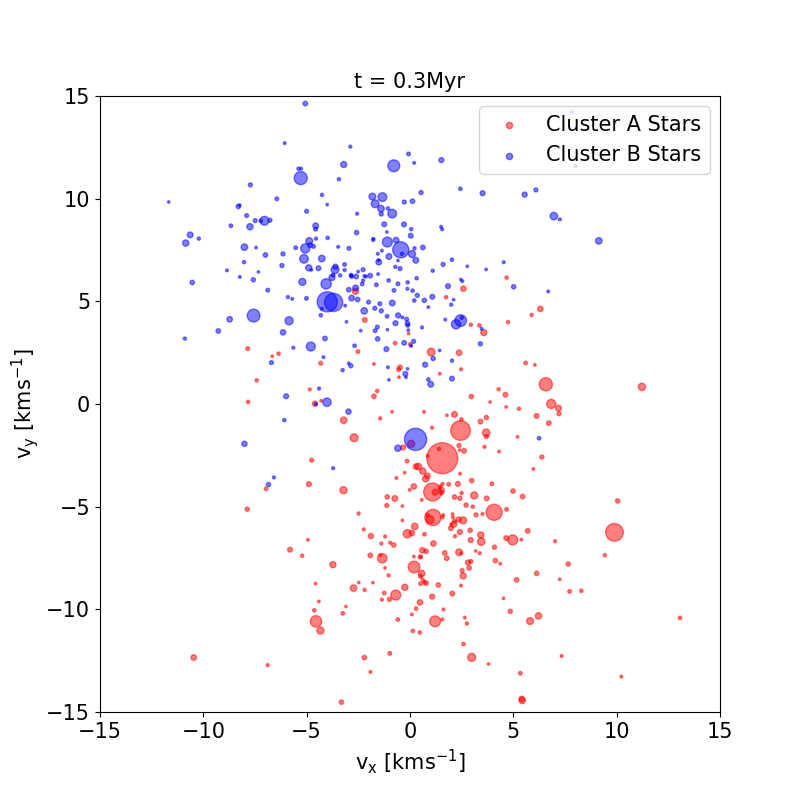}
    \includegraphics[scale=0.27]{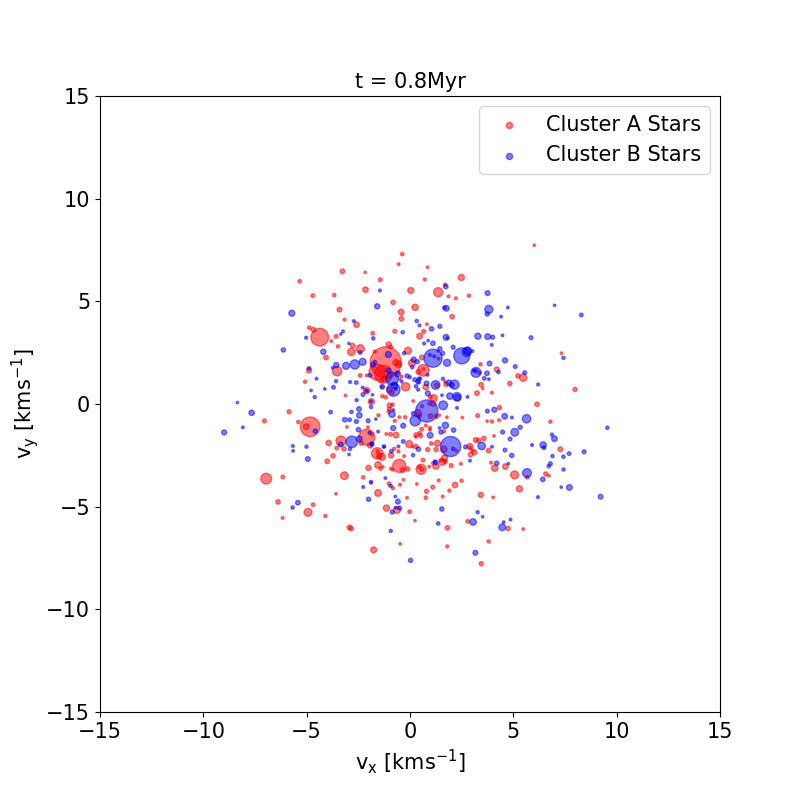}
    \caption{Velocities of stars originally belonging to clusters A (red) and B (blue) around the centre of mass velocity of cluster AB at two snapshots in the \texttt{region1} simulation. The size of the circles scales with the mass of the star.}
    \label{fig:vel_dist1}
\end{figure}

\begin{figure*}
    \centering
    \hspace{25pt}
    \includegraphics[scale=0.4]{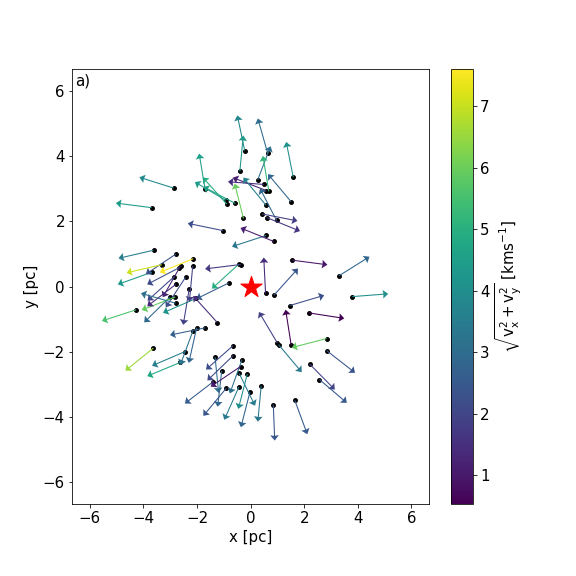}
    \includegraphics[scale=0.4]{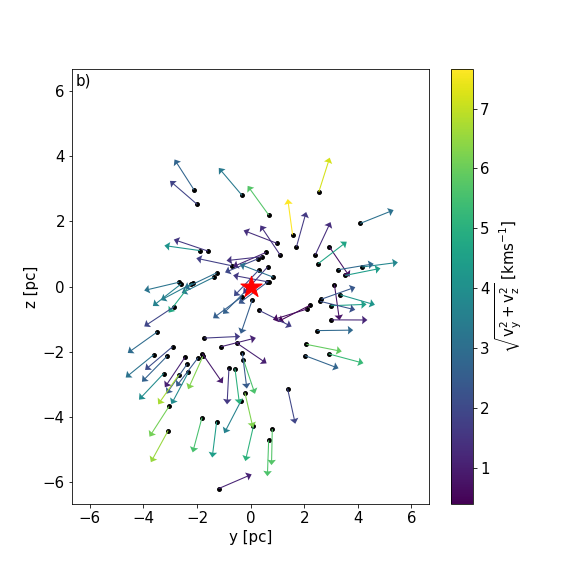}
    \includegraphics[scale=0.32]{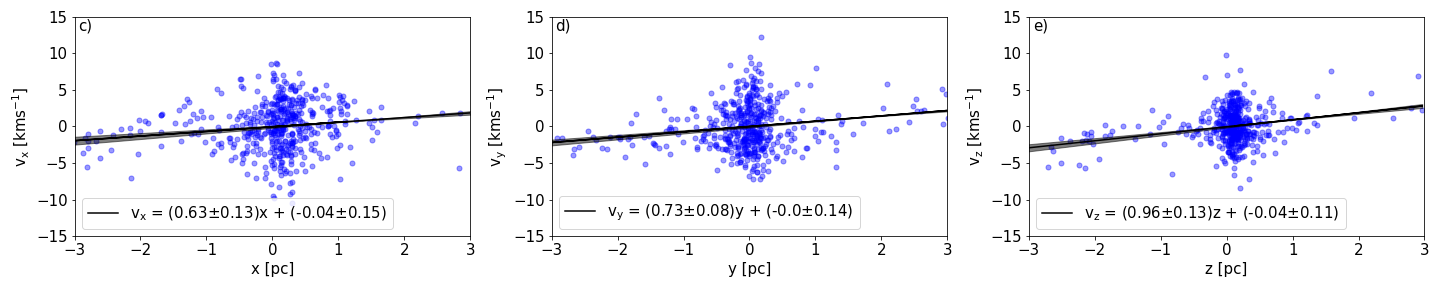}
    \caption{Expansion of cluster AB in the \texttt{region1} simulation at the end of the simulation. All positions and velocities in this figure are about the centre of mass and centre of mass velocity respectively. a), b): Location of stars (black points) beyond the 90\% Lagrangian radius at the end of the simulation in x-y and y-z planes. Arrows are unit vectors which show direction of the velocity vectors in the corresponding plane. The colour of the arrows corresponds to the magnitude of the velocity vectors. The red star indicates the location of the centre of mass of cluster AB. c), d), e): Position of each star along x, y, and z plotted against velocity of the star in the same direction. Black lines show the line of best fit with shaded regions showing one sigma of the fit calculated through bootstrapping 10$^5$ times.}
    \label{fig:spur_quiver}
\end{figure*}

Figure \ref{fig:sigma1} shows that, around the time of the merger of cluster A and B at 0.1-0.4Myr, the velocity dispersions of the stars involved in the merger increase by more than a factor of two. Because the merger of clusters A and B results in a large and rapid change in the potential felt by the stars involved, this can be described as a violent relaxation process \citep{tvr}. We estimate the violent relaxation timescale to be $\approx$0.4Myr for this merger. This increase lasts for approximately one violent relaxation time, after which the velocity dispersions stabilize. As well, we find that the velocity dispersion of the stars belonging to cluster AB are independent of their mass. This confirms that violent relaxation is occurring in the resultant cluster \citep{javier}. 

We use the virial parameter (defined as $\alpha=2K/|U|$ where $K$ is the kinetic energy of the stars and gas bound to the cluster, and $U$ is the potential energy of the stars and gas bound to the cluster) to analyze whether cluster AB is stable after the merger. We calculate a virial parameter of $\alpha = 0.38$ at the end of the simulation implying that cluster AB is bound even after the increase in the velocity dispersion of the stellar component. This is consistent with the low fraction of unbound stars after the merger.

Next, we consider whether the stars that make up the merged cluster AB carry with them any memory of their velocity space structure before the merger. We show snapshots of the x and y velocities of the cluster A and B stars subtracted by the centre of mass velocity of cluster AB in Figure \ref{fig:vel_dist1} at different times in the simulation. The top panel shows a snapshot of the stars in velocity space during the merger, and the bottom panel shows a snapshot corresponding to one violent relaxation time after the merger has finished. During the merger, the two groups of stars are most distinguishable. This is quickly erased after $\approx$1 violent relaxation time. They stay mixed until the end of our simulation. We find the same trend in the distribution of velocities along the y and z axes. We conclude that the violent relaxation process has removed velocity signatures that were unique to the stars belonging to the clusters involved in this merger.

\subsubsection{Expansion and Rotation}
\label{sec:exp_rot_spur}
After an initial growth, the 10\%, 50\%, and 75\% Lagrangian radii stay roughly constant after the merger. The 90\% Lagrangian radius increases for the entirety of the simulation after the merger implying that the majority of the expansion of cluster AB after the merger is beyond the 75\% Lagrangian radius. Figure \ref{fig:spur_quiver} shows an analysis of the expansion of cluster AB. We plot different measures of the expansion around the centre of cluster AB at the end of the \texttt{region1} simulation. The top panels show the orientation of the velocity vectors of the stars beyond the 90\% Lagrangian radius of cluster AB in the x-y and y-z planes. From these panels, we see that the expansion is not isotropic at the end of the simulation in either plane. 

\begin{figure}
    \centering
    \includegraphics[scale=0.4]{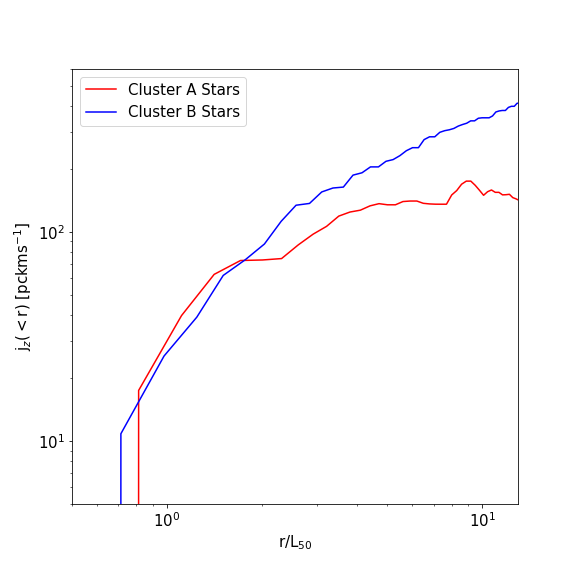}
    \caption{Cumulative z-component specific angular momentum of cluster AB stars. The red line shows stars that originally belonged to cluster A, and the blue line shows stars that originally belonged to cluster B.}
    \label{fig:cum_ang_mom_spur}
\end{figure}

We calculate the rate of expansion throughout the merger using the method outlined in \citet{wright}. An example of this process applied to cluster AB at the end of the simulation can be seen in the bottom panels of Figure \ref{fig:spur_quiver}. We first plot the position of the stars along a given axis against the velocity of the stars along that same axis (e.g. x vs v$_{\mathrm{x}}$). Next, we fit a line to the data (black lines in panels c), d) and e) of Figure \ref{fig:spur_quiver}). The slope of that line gives us a measure of the rate of the expansion. A positive slope refers to expansion while a negative slope refers to contraction. We do this for all times after the merger and find that the rate of the expansion of the stellar component of cluster AB is not constant over time. During the merger, the expansion rate is negative, indicating that the stellar component is contracting as the stars of cluster A and B move towards each other. After the merger, the expansion rates in x, y and z follow a similar trend by becoming positive and decreasing slowly until the end of the simulation. This implies that the expansion is slowing down after the merger along all three axes. The expansion rates at the end of the simulation can be seen as the slopes of the lines of best fit in the bottom panels of Figure \ref{fig:spur_quiver}. These values are in the similar range to those found for open clusters along the plane of the sky (\citealt{wright_2}).

Stars belonging to cluster AB gain angular momentum about the centre of mass of the stellar component of cluster AB from the merger of cluster A and B. The z-component of angular momentum peaks during the merger due to the counter-clockwise rotation of the stars in the x-y plane. It decreases slowly for the remainder of the simulation. The majority of the specific angular momentum inherited from this merger is in the inner regions (within $\approx 3$L$_{\mathrm{50}}$) of the cluster. We show an example of this in Figure \ref{fig:cum_ang_mom_spur}. Such a trend with rotation amplitude has been observed in intermediate age star clusters (e.g. \citealt{ngc_1846}). As well, we find a difference in the distribution of specific angular momentum for the stars originally belonging to cluster A and those originally belonging to cluster B in the outer regions of cluster AB similar to the result found by \citet{lahen}. Lastly, we find that not all stars have the same direction of angular momentum. We discuss this more in Section \ref{sec:discussion}.

\begin{figure*}

    \centering
    
    \includegraphics[scale=0.35]{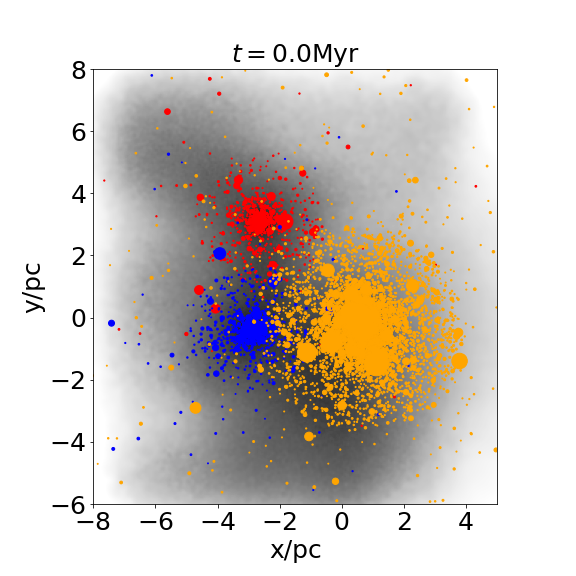}
    \includegraphics[scale=0.35]{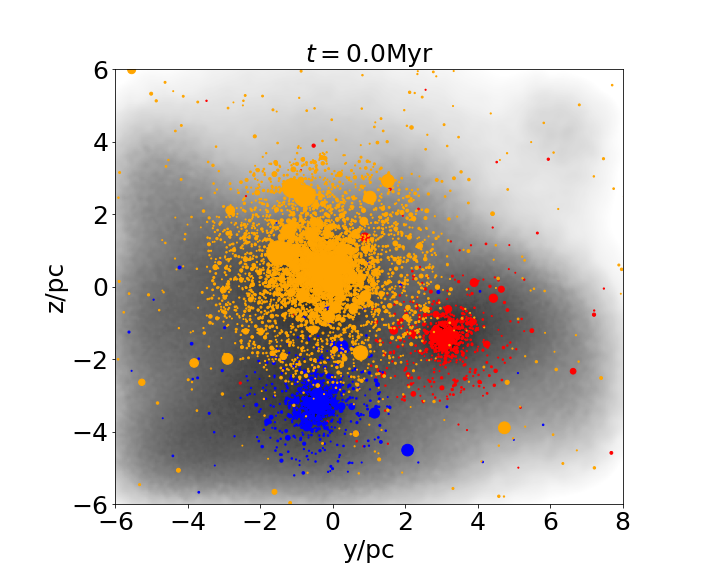}
    
    \includegraphics[scale=0.35]{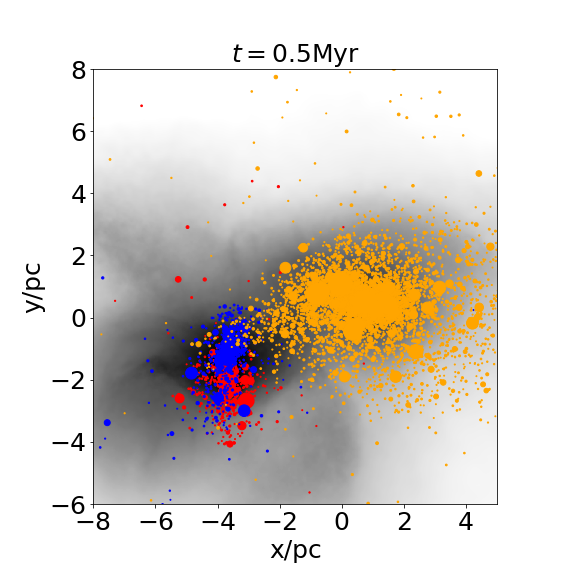}
    \includegraphics[scale=0.35]{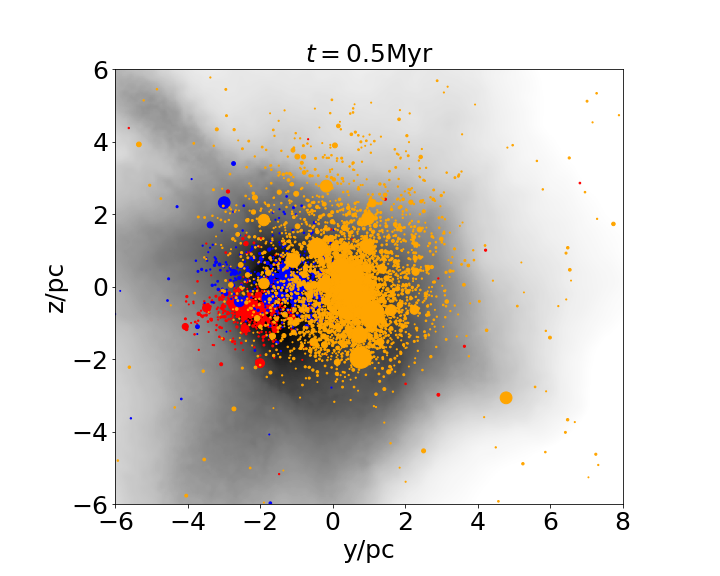}
    
    \includegraphics[scale=0.35]{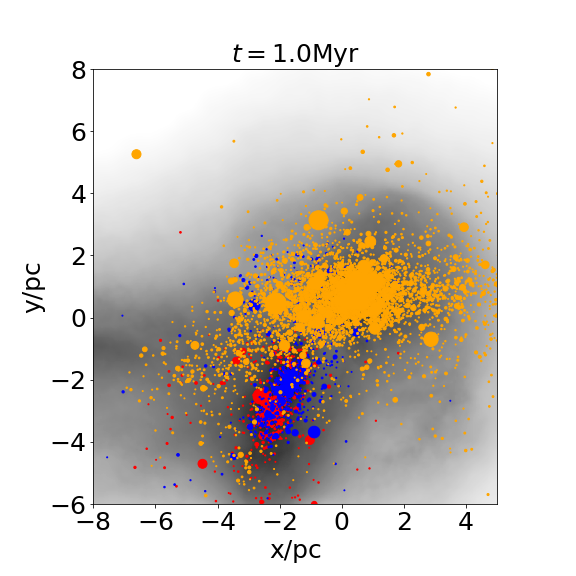}
    \includegraphics[scale=0.35]{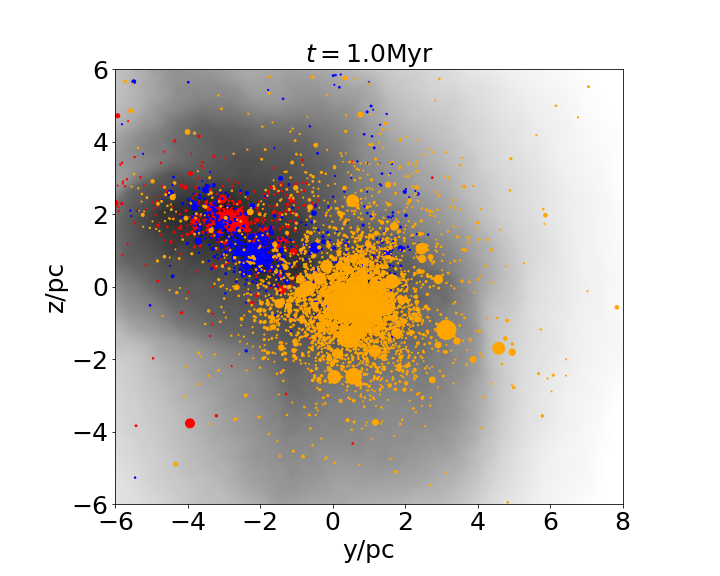}
    
    \caption{Same as Figure \ref{fig:spur_snaps} but for the \texttt{region2} simulation. The lowest and highest mass stars in this simulation are 0.15M$_\odot$ and 98.34M$_\odot$ respectively.}
    \label{fig:r3_snaps}
\end{figure*}

\subsection{Region 2}
We now discuss the evolution of the stellar components of the clusters present in the \texttt{region2} simulation outlined in Table \ref{tab:regions}. We show snapshots of the distribution of stars and gas present in a portion of this simulation in Figure \ref{fig:r3_snaps}. The circles are colour-coded depending on the cluster to which they initially belong with red, blue, and orange corresponding to clusters A, B, and C respectively. Initially, clusters A and B are travelling towards one another in the x-y plane.

At t$\approx$ 0.4Myr, cluster A has begun merging with cluster B. The impact parameter of this merger is 1.9L$_{\mathrm{50,B}}$ where L$_{\mathrm{50,B}}$ is the 50\% Lagrangian radius of cluster B. At this point stars from cluster C begin moving towards the merged cluster due to the increased potential felt. As a result, the merged cluster accretes an extra $\approx 300$M$_{\odot}$ from cluster C by the end of the simulation. We call this final resultant cluster ``cluster ABc''. Because only $\approx 1\%$ of the stellar mass becomes unbound by the end of the simulation, the total mass of the cluster ABc increases beyond the sum of the masses of cluster A and B.

\subsubsection{Velocity Space Distribution}
\label{sec:vel_space_r3}
Similarly to the stars in the \texttt{region1} simulation, the merger of cluster A and cluster B results in a sharp and temporary increase in the velocity dispersion of the stars involved during the merger. The virial parameter of the merged cluster is $\alpha = 0.51$ by the end of the simulation indicating that cluster ABc is bound.

The stars originally belonging to clusters A and B are mixed in velocity space by the end of the simulation similarly to the \texttt{region1} simulation. The stars accreted onto the merged cluster from cluster C are mixed in velocity space with the rest of the stars belonging to cluster ABc by the end of the simulation. It would therefore be difficult to determine whether stars in this cluster were accreted from a nearby cluster using their distribution in velocity space.

\begin{figure}
    \centering
    \includegraphics[scale=0.4]{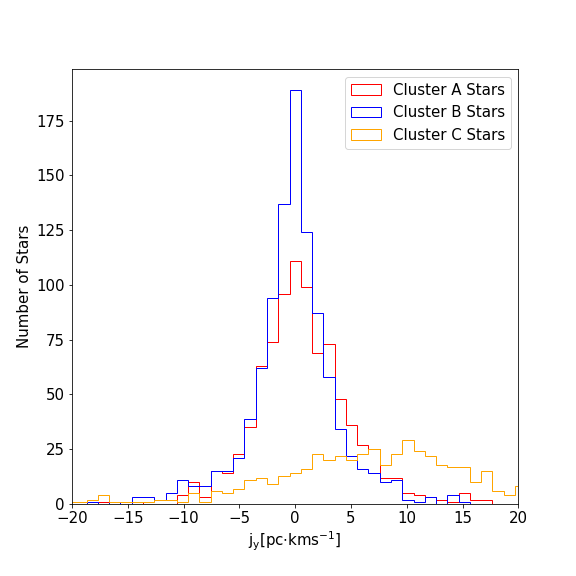}
    \caption{Distribution of y-component specific angular momentum of all stars in cluster ABc at the end of the \texttt{region2} simulation. Red, blue, and orange lines show stars which originally belonged to cluster A, B, and C respectively.}
    \label{fig:jy_region3}
\end{figure}

\begin{figure*}

    \centering
    
    \includegraphics[scale=0.35]{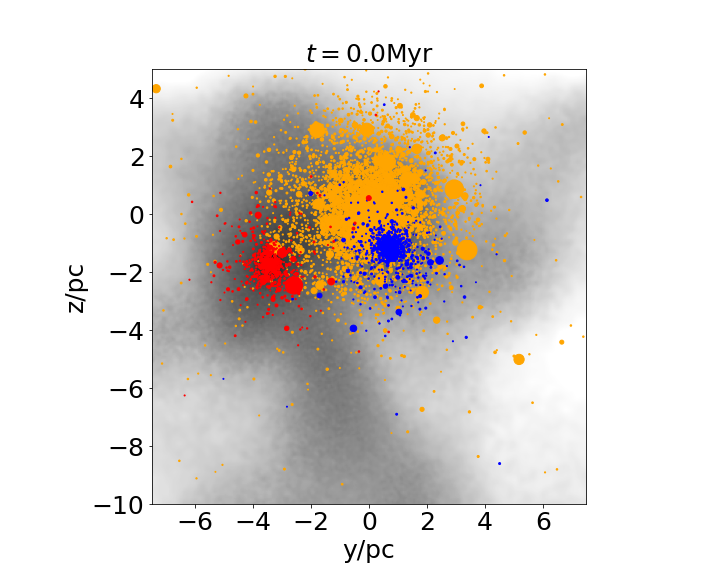}
    \includegraphics[scale=0.35]{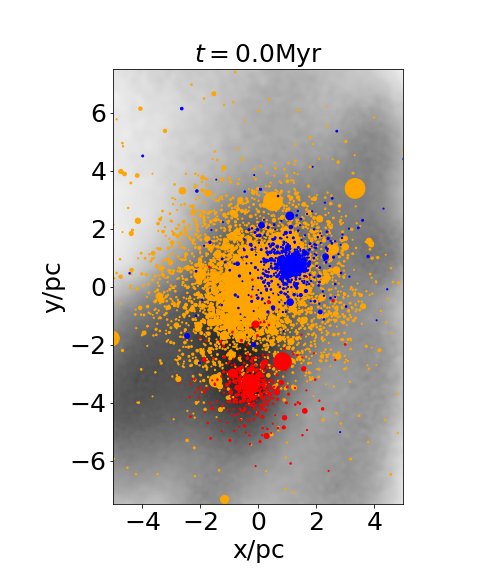}
    
    \includegraphics[scale=0.35]{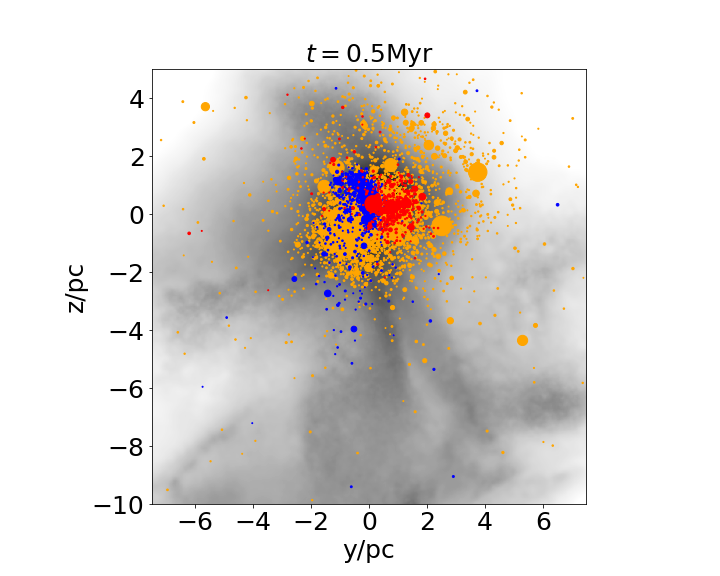}
    \includegraphics[scale=0.35]{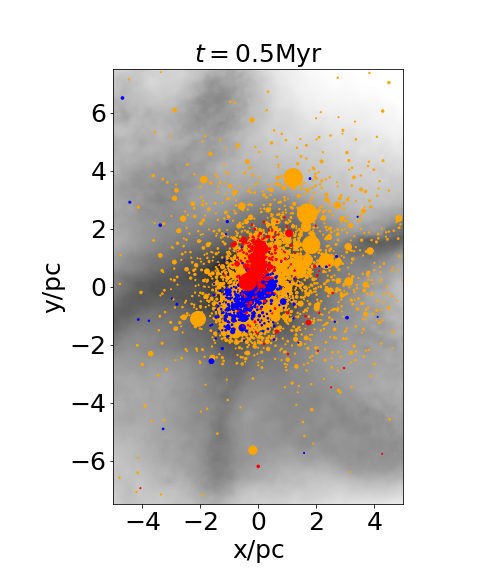}
    
    \includegraphics[scale=0.35]{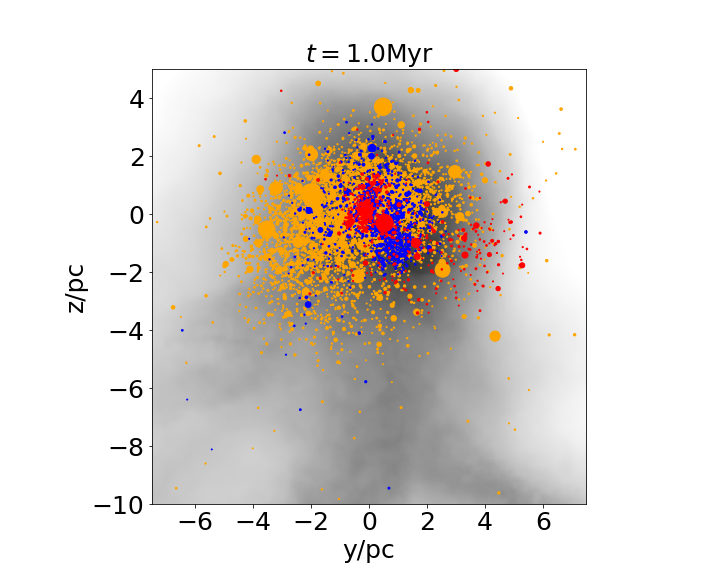}
    \includegraphics[scale=0.35]{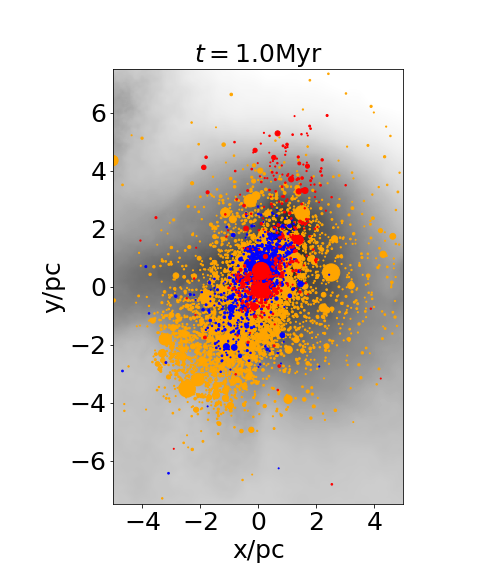}
    
    \caption{Same as Figure \ref{fig:spur_snaps} but for the \texttt{region3} simulation. The lowest and highest mass stars in this simulation are 0.15M$_\odot$ and 74.59M$_\odot$ respectively.}
    \label{fig:jelly_snaps}
\end{figure*}

\subsubsection{Expansion and Rotation}
The 50\%, 75\% and 90\% Lagrangian radii of cluster ABc all increase after the merger. This increase stops by the end of the simulation. We find a positive expansion rate in the stellar component of cluster ABc along the x and y axes at the end of the simulation. There is no clear signal of expansion along the z-axis. The values of the expansion rates are 0.19 $\pm$ 0.02, and 0.14 $\pm$ 0.04 kms$^{-1}$pc$^{-1}$ along the x and y axes respectively at the end of the simulation. These expansion rates are much lower than those found along all axes in cluster AB in the \texttt{region1} simulation. This could be due to the strong potential which results from the high mass in stars and gas present in the \texttt{region2} simulation. The small expansion that is present around cluster ABc in this simulation is anisotropic similar to that found in cluster AB in the \texttt{region1} simulation.

We see small changes to the shape and size of cluster C as a result of the stripped stars and the gravitational effect of the nearby cluster ABc. The presence of contraction or expansion along only two of the three axes as we see in clusters ABc and C in this simulation implies that the shape of the stellar component of the clusters has changed to become less spherical. This is consistent with results presented in \citet{claude_2023} who find that cluster mergers can cause clusters to become more elliptical.

There is angular momentum in the stellar component of cluster ABc. Stars that are now part of cluster ABc but originally belonged to clusters A, B, or C all share a similar distribution of x-component specific angular momentum. Along both the y and z axes, we find that most of the angular momentum comes from the accreted stars. We show an example of this for the y-component of the specific angular momentum in Figure \ref{fig:jy_region3}. Stars that originally belonged to cluster C have more total y-component specific angular momentum than stars that originally belonged to clusters A or B. Accretion from surrounding clusters can be important in inducing angular momentum onto the clusters that are accreting. Because of the high specific angular momentum of the accreted stars in the outskirts of cluster ABc, the distribution of angular momentum does not have a concentration within 3L$_{\mathrm{50}}$ as it did in cluster AB from the \texttt{region1} simulation.



\subsection{Region3}
We now discuss the evolution of the stellar components of the clusters in the \texttt{region3} simulation. We show snapshots of a portion of this simulation in Figure \ref{fig:jelly_snaps}. In this simulation, both the red and blue coloured clusters (clusters A and B respectively), are initially moving clockwise around the much more massive orange coloured cluster (cluster C) in the y-z and x-y planes. After $\approx 0.2$Myr, both cluster A and cluster B have merged with cluster C. We call the final resultant cluster ``cluster ABC''. The impact parameters of the mergers of cluster A and cluster B with cluster C are 1.1L$_{\mathrm{50,C}}$ and 0.7L$_{\mathrm{50,C}}$ respectively where L$_{\mathrm{50,C}}$ is the 50\% Lagrangian radius of cluster C.

The mergers result in $<$0.05\% of the total stellar mass becoming unbound from cluster ABC by the end of the simulation.

\subsubsection{Velocity Space Distribution}
Similarly to the mergers in the \texttt{region1} and \texttt{region2} simulations, we find that the merger of clusters A and B with cluster C results in a sharp, but temporary, increase in the velocity dispersion of the stars. After $\approx 0.5$Myr, the velocity dispersion decreases down to $\approx 8$kms$^{-1}$. The virial parameter of cluster ABC at the end of the simulation is $\alpha = 0.74$ indicating that the cluster is bound. After the mergers, the stars from all three clusters have mixed in velocity space similarly to the \texttt{region1} and \texttt{region2} simulations.

\subsubsection{Expansion and Rotation}

\begin{figure}
    \centering
    \includegraphics[scale=0.33]{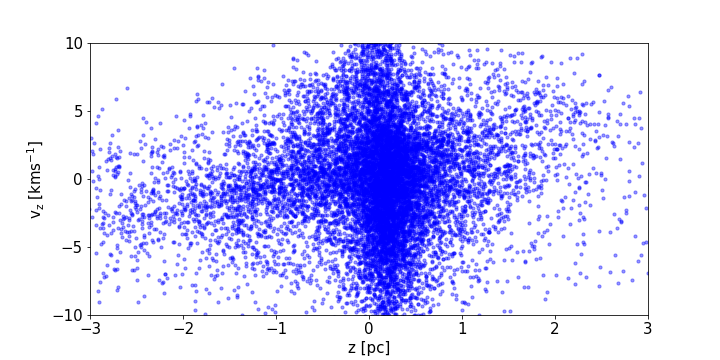}
    \includegraphics[scale=0.33]{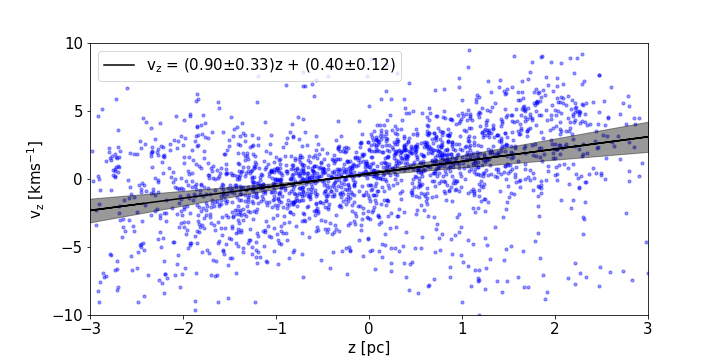}
    \caption{Position-velocity distribution of stars in cluster ABC from the \texttt{region3} simulation along the z-axis at the end of the simulation. Top panel shows all stars, and bottom panel shows all stars between the 75\% and 90\% Lagrangian radii of the cluster. The black line in the bottom panel shows the line of best fit with the shaded regions showing one sigma of the fit calculated through bootstrapping 10$^5$ times.}
    \label{fig:expansion_jelly}
\end{figure}

For a short time after the mergers of clusters A and B with cluster C in the \texttt{region3} simulation, we see signatures of contraction. The 90\% and 75\% Lagrangian radii are decreasing for $\approx$0.6Myr after the merger. After this time, they are increasing until the end of the simulation. Below the 75\% Lagrangian radii, we do not see signatures of expansion or contraction indicating that the expansion and contraction are mostly present in the outer regions of the cluster.

This can also be seen in the distribution of the cluster ABC stars in position-velocity space. An example is shown in Figure \ref{fig:expansion_jelly}. In the top panel, we show the distribution of all stars in position-velocity space along the z axis at the end of the simulation. From this, we see no clear signature of expansion. The bottom panel of Figure \ref{fig:expansion_jelly} shows the position-velocity distribution along the z axis of stars between 75\% and 90\% Lagrangian radii of cluster ABC at the same time. Here, we see a higher linear correlation implying that the expansion signature is stronger when considering these stars. When considering stars beyond the 90\% Lagrangian radius, we find that the expansion signature decreases. The same thing is true when we consider stars within the 75\% Lagrangian radius. We find similar trends when looking along the x and y axes. As well, we find that the expansion is much more anisotropic along the x and y axes than it is along the z-axis with most of the expansion taking place along the negative x and y axes.

There is negative x-component angular momentum induced in cluster ABC stars after the mergers. As the stars belonging to cluster A and B rotate around cluster C, they impart angular momentum on the initially non-rotating stellar component of cluster C. As well, we find that not all stars are rotating clockwise consistent with the merger direction. This is similar to the previous simulations in this section. The angular momentum of cluster ABC is concentrated within $\approx 3$L$_{50}$.

\section{Zoom-In Regions: Gas}

\label{sec:gas}

In all three zoom-in simulations, we find that the amount of unbound gas mass is negligible throughout the simulations. In all cases, the amount of unbound gas is less than that from the merger simulations presented in Paper I and in Section \ref{sec:stars} likely because of the background gas distribution providing pressure to keep gas bound to the cluster (see Paper II).

\begin{figure}
    \centering
    \includegraphics[scale=0.33]{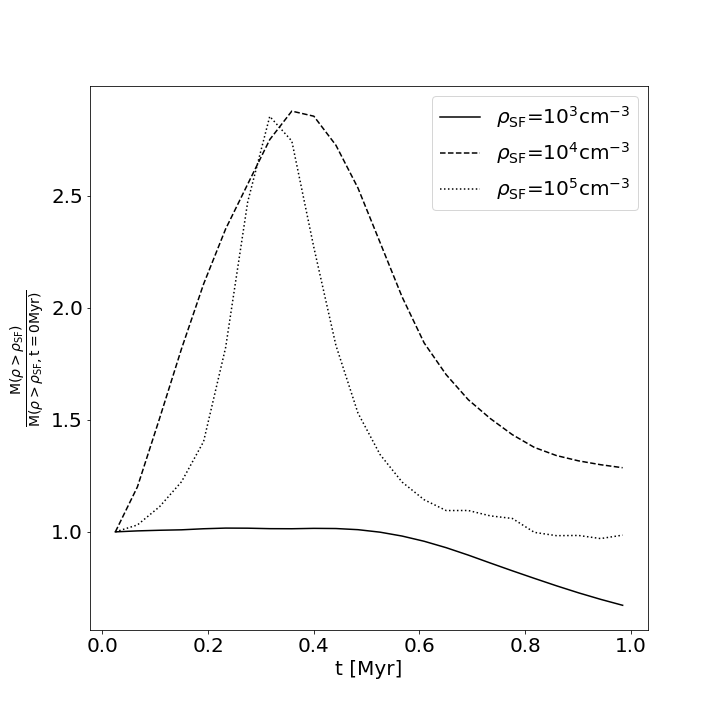}
    \caption{Change in the total mass of gas above 10$^3$cm$^{-3}$ (solid), 10$^4$cm$^{-3}$ (dashed) and 10$^5$cm$^{-3}$ (dotted) for all gas in the \texttt{region1} simulation.}
    \label{fig:gas_above_SF_spur}
\end{figure}

We find that the amount of gas with densities above 10$^4$ and 10$^5$cm$^{-3}$ increases from the merger processes in all three zoom-in simulations. These increases last for $\approx$2 free-fall times for the gas at that density threshold. After this, they decrease back down to their original values. The amount of gas above 10$^3$cm$^{-3}$ stays constant for most of the simulation for all three simulations. We show an example of this behaviour in the \texttt{region1} simulation in Figure \ref{fig:gas_above_SF_spur}. In our simulations that did not include a background gas component, we found that the amount of dense gas decreases below its original value by the end of the simulation. This illustrates another important property of the background gas: its presence may help in promoting star formation through cluster mergers.

At the beginning of the mergers in all three simulations, we see a bridge of gas with density above 10$^4$cm$^{-3}$ connecting the cores of the clusters involved in the merger. We expect that the newly formed stars will mix with the rest of the stars in the cluster one free-fall time later because the bridge of star forming gas connecting the two cores has shrunk.

In velocity space, at the beginning of the merger, we find that the distributions of stars and gas look different from each other. In the \texttt{region1} and \texttt{region2} simulations, the distributions of the stars in velocity space are mostly flat. The gas however, is peaked around the centre of mass velocity of the merged cluster in each simulation. Such a strong distinction between the distributions of the stars and gas in velocity space is not present in the \texttt{region3} simulation. We discuss the implications of this in Section \ref{sec:discussion}.

Converse to the velocity space distribution of stars and gas in our simulations, the distributions of specific angular momentum are similar between the stars and gas after the mergers in each simulation. The peaks in the specific angular momentum distributions of the stars overlap the peaks in the specific angular momentum distributions of the gas in the \texttt{region1} and \texttt{region2} simulations. In the \texttt{region3} simulations, differences between the stellar and gas distribution in specific angular momentum space are not strong.

\section{Summary and Discussion}
\label{sec:discussion}

We have analyzed the role played by the impact parameter in the merging of two gas-rich star clusters, and how environment affects the dynamics of merging clusters. We have done this through a combination of off-axis merger simulations that do not include a background gas component, and simulations that zoom into a previously run GMC simulation from \citet{Howard2018} (H18). The impact parameter of a merger can have a small role in determining the amount of bound stars and gas the resultant cluster can maintain. High impact parameters can also reduce the collisional velocity necessary to produce a non-monolithic resultant cluster. However, our zoom-in simulations show that background gas is a very important component in driving the motions of clusters as they merge. It also helps keep clusters monolithic. The dynamics of the stars are affected by the merger process. Velocity dispersions of the stars involved in the mergers increase as a result of the merger, and the stars completely mix in velocity space shortly after the merger finishes. Mergers also induce expansion and angular momentum in the stellar component of the resultant cluster. This expansion is anisotropic in all zoom-in simulations studied in this work. The angular momentum of the stellar component is concentrated in the inner regions of the resultant cluster unless that cluster accretes stars from a nearby cluster. Mergers increase the amount of potentially star-forming dense gas present in the cluster when a background gas distribution is included.

We have found that the presence of background gas is extremely important in determining the motion of clusters that are embedded inside GMCs. Through feedback and star formation, the mass in gas inside a star forming GMC decreases while the mass in stars increases (e.g. \citealt{li_2019}, \citealt{fujii_sirius_3}). In the context of our simulations, this implies that mergers between clusters inside GMCs may become less monolithic with time.

The sink particle prescription outlined in \citet{sink} does not allow for sinks to lose mass. From our results, we find that mergers of clusters inside the GMC environment result in a negligible amount of unbound material implying that this component of the sink particle prescription holds in our simulations. Sinks can only gain mass through mergers with other sinks and gas accretion. This results in a discrepancy between our \texttt{region2} simulation, and its sink particle analog because of the accretion of stars from cluster C onto the merged cluster. The spatial distribution of stars in a cluster should be considered in sink particle prescriptions.

The expansion of the stellar component we see in our clusters is different from expansion observed in star clusters after they have removed their gas component through feedback. For embedded clusters, expansion is expected as gas is being removed from the cluster through feedback effects because of the drastic decrease in the potential of the system (e.g. \citealt{pelupessy}, \citealt{expansion_pfalzner}, \citealt{farias}, \citealt{arunima}). In particular, \citet{pelupessy} show that through gas expulsion the 50, 75, and 90\% Lagrangian radii are all expected to grow well after the initial expansion from gas loss. In our \texttt{region1} and \texttt{region2} simulations, the 50\% and 75\% Lagrangian radii of the merged clusters plateau after some initial growth likely because the potential from the background gas is still present. As well, the expansion found in \citet{pelupessy} was accompanied by a drastic loss in bound stellar mass which we do not see in any of our simulations.

We also see that the expansion of all of our merged clusters is anisotropic. While the observations performed by \citet{wright_2} have found anisotropic expansion around embedded young clusters, most observations are limited to clusters that have dispersed their natal gas clouds. It is therefore important to test whether the anisotropy of the expansion found in our simulations remains as the cluster expands its surrounding gas component.

The simulations from H18 show that clusters merge many times during their embedded phase. These mergers happened on average once every $\approx$0.4Myr. As we have seen from our simulations, not all stars have angular momentum in the same direction around the centre of mass of the cluster. This is expected to persist as clusters build up and merge throughout the GMCs life.

We can predict the number of stars expected to form from the dense gas present in our simulations by considering a star formation efficiency per free-fall time of $\epsilon_{\mathrm{ff}} = 3$\% that matches recent observations of star forming molecular clouds (\citealt{eff}). Using this value we find that the stars that form from the dense gas in the \texttt{region1} and \texttt{region2} simulations will affect the overall distribution of stars in velocity space making it more peaked around the centre of mass velocity of the resultant cluster. Therefore, clusters may be able to develop stellar age gradients associated with kinematic subgroups as as they evolve. This is similar to those observed within subgroups in older OB associations (e.g. \citealt{pecaut_sco_cent}, \citealt{US_kin_subgroups}). In the \texttt{region3} simulation, because the number of stars belonging to cluster ABC is so large, we predict that the newly formed stars will have very little effect on the overall distribution in position, velocity, or angular momentum space. This implies that mergers that involve clusters consisting of a high mass stellar component may not create easily detectable gradients in age or kinematic subgroups. Because of the high mass in gas present in GMCs early on in their evolution, we expect mergers that take place early on in the GMCs life to lead to more noticeable kinematic and age distinctions in stars. To fully understand observations of phase space subgroups in clusters and associations, we must consider multiple subcluster mergers, and employ realistic star formation and feedback prescriptions.

\begin{acknowledgments}
The authors thank Claude Cournoyer-Cloutier, Marta Reina-Campos, Veronika Dornan, and Raven Westlake for helpful discussions. The authors also thank the referee for careful reading and constructive comments. AS is supported by the Natural Sciences and Engineering Research Council of Canada. This research was enabled in part by suport provided by Compute Ontario (https://www.computeontario.ca) and Compute Canada
(http://www.computecanada.ca).
\end{acknowledgments}

%

\vspace{5mm}


\software{astropy \citep{2013A&A...558A..33A,2018AJ....156..123A},  
          numpy \citep{numpy},
          matplotlib \citep{matplotlib},
          pynbody \citep{pynbody},
          scipy \citep{2020SciPy-NMeth}
          }





\bibliography{sample631}{}
\bibliographystyle{aasjournal}



\end{document}